\input harvmac
\let\includefigures=\iftrue
\let\useblackboard==\iftrue
\newfam\black

\includefigures
\message{If you do not have epsf.tex (to include figures),}
\message{change the option at the top of the tex file.}
\input epsf
\def\figin{\epsfcheck\figin}\def\figins{\epsfcheck\figins}
\def\epsfcheck{\ifx\epsfbox\UnDeFiNeD
\message{(NO epsf.tex, FIGURES WILL BE IGNORED)}
\gdef\figin##1{\vskip2in}\gdef\figins##1{\hskip.5in}
\else\message{(FIGURES WILL BE INCLUDED)}%
\gdef\figin##1{##1}\gdef\figins##1{##1}\fi}
\def\DefWarn#1{}
\def\figinsert{\goodbreak\midinsert}
\def\ifig#1#2#3{\DefWarn#1\xdef#1{fig.~\the\figno}
\writedef{#1\leftbracket fig.\noexpand~\the\figno}%
\figinsert\figin{\centerline{#3}}\medskip\centerline{\vbox{
\baselineskip12pt\advance\hsize by -1truein
\noindent\footnotefont{\bf Fig.~\the\figno:} #2}}
\endinsert\global\advance\figno by1}
\else
\def\ifig#1#2#3{\xdef#1{fig.~\the\figno}
\writedef{#1\leftbracket fig.\noexpand~\the\figno}%
\global\advance\figno by1} \fi

\def\id{{1 \kern-.28em {\rm l}}}

\def\O{{\cal O}}

\def\K3{{\bf K3}}
\def\journal#1&#2(#3){\unskip, \sl #1\ \bf #2 \rm(19#3) }
\def\andjournal#1&#2(#3){\sl #1~\bf #2 \rm (19#3) }

\def\bar{\overline}
\def\hat{\widehat}
\def\ie{{\it i.e.}}
\def\eg{{\it e.g.}}

\def\frac#1#2{{#1\over#2}}

\def\half{\frac12}

\def\inbar{\,\vrule height1.5ex width.4pt depth0pt}
\def\IC{\relax\hbox{$\inbar\kern-.3em{\rm C}$}}
\def\IR{\relax{\rm I\kern-.18em R}}
\def\IP{\relax{\rm I\kern-.18em P}}

%
%

%
\catcode`\@=11
\def\slash#1{\mathord{\mathpalette\c@ncel{#1}}}
\overfullrule=0pt

\def\LL{{\cal L}}
\def\MM{{\cal M}}
\def\NN{{\cal N}}
\def\OO{{\cal O}}

\def\underrel#1\over#2{\mathrel{\mathop{\kern\z@#1}\limits_{#2}}}

\catcode`\@=12


%

\def\exp{{\rm exp}}

\def\O{{\cal O}}

\def\ie{{\it i.e.}}
\def\eg{{\it e.g.}}

\lref\TeschnerFT{
  J.~Teschner,
  ``On structure constants and fusion rules in the SL(2,C) / SU(2) WZNW model,''
Nucl.\ Phys.\ B {\bf 546}, 390 (1999).
[hep-th/9712256].
}

\lref\IsraelIR{
  D.~Israel, C.~Kounnas, A.~Pakman and J.~Troost,
  ``The Partition function of the supersymmetric two-dimensional black hole and little string theory,''
JHEP {\bf 0406}, 033 (2004).
[hep-th/0403237].
}

\lref\MaldacenaHW{
  J.~M.~Maldacena and H.~Ooguri,
  ``Strings in AdS(3) and SL(2,R) WZW model 1.: The Spectrum,''
J.\ Math.\ Phys.\  {\bf 42}, 2929 (2001).
[hep-th/0001053].
}

\lref\ZamolodchikovCE{
  A.~B.~Zamolodchikov,
  ``Expectation value of composite field T anti-T in two-dimensional quantum field theory,''
[hep-th/0401146].
}

\lref\SmirnovLQW{
  F.~A.~Smirnov and A.~B.~Zamolodchikov,
  ``On space of integrable quantum field theories,''
[arXiv:1608.05499 [hep-th]].
}

\lref\CavagliaODA{
  A.~Cavagliˆ, S.~Negro, I.~M.~SzŽcsŽnyi and R.~Tateo,
  ``$T \bar{T}$-deformed 2D Quantum Field Theories,''
[arXiv:1608.05534 [hep-th]].
}

\lref\MooreGA{
  G.~W.~Moore,
  ``Gravitational phase transitions and the Sine-Gordon model,''
[hep-th/9203061].
}

\lref\HsuCM{
  E.~Hsu and D.~Kutasov,
  ``The Gravitational Sine-Gordon model,''
Nucl.\ Phys.\ B {\bf 396}, 693 (1993).
[hep-th/9212023].
}

\lref\GiveonCMA{
  A.~Giveon, N.~Itzhaki and D.~Kutasov,
  ``Stringy Horizons,''
JHEP {\bf 1506}, 064 (2015).
[arXiv:1502.03633 [hep-th]].
}

\lref\GiveonNIE{
  A.~Giveon, N.~Itzhaki and D.~Kutasov,
  ``$T\bar T$ and LST,''
[arXiv:1701.05576 [hep-th]].
}

\lref\PolchinskiRQ{
  J.~Polchinski,
  ``String theory. Vol. 1: An introduction to the bosonic string,''
}

\lref\GiveonCGS{
  A.~Giveon and D.~Kutasov,
  ``Supersymmetric Renyi entropy in CFT$_{2}$ and AdS$_{3}$,''
JHEP {\bf 1601}, 042 (2016).
[arXiv:1510.08872 [hep-th]].
}

\lref\KutasovXU{
  D.~Kutasov and N.~Seiberg,
  ``More comments on string theory on AdS(3),''
JHEP {\bf 9904}, 008 (1999).
[hep-th/9903219].
}

\lref\GiveonNS{
  A.~Giveon, D.~Kutasov and N.~Seiberg,
  ``Comments on string theory on AdS(3),''
Adv.\ Theor.\ Math.\ Phys.\  {\bf 2}, 733 (1998).
[hep-th/9806194].
}

\lref\GiveonUP{
  A.~Giveon and D.~Kutasov,
  ``Notes on AdS(3),''
Nucl.\ Phys.\ B {\bf 621}, 303 (2002).
[hep-th/0106004].
}

\lref\MaldacenaKM{
  J.~M.~Maldacena and H.~Ooguri,
  ``Strings in AdS(3) and the SL(2,R) WZW model. Part 3. Correlation functions,''
Phys.\ Rev.\ D {\bf 65}, 106006 (2002).
[hep-th/0111180].
}

\lref\ArgurioTB{
  R.~Argurio, A.~Giveon and A.~Shomer,
  ``Superstrings on AdS(3) and symmetric products,''
JHEP {\bf 0012}, 003 (2000).
[hep-th/0009242].
}

\lref\GiveonMI{
  A.~Giveon, D.~Kutasov, E.~Rabinovici and A.~Sever,
  ``Phases of quantum gravity in AdS(3) and linear dilaton backgrounds,''
Nucl.\ Phys.\ B {\bf 719}, 3 (2005).
[hep-th/0503121].
}

\lref\GiveonJG{
  A.~Giveon and M.~Rocek,
  ``Supersymmetric string vacua on AdS(3) x N,''
JHEP {\bf 9904}, 019 (1999).
[hep-th/9904024].
}

\lref\PeetWN{
  A.~W.~Peet and J.~Polchinski,
  ``UV / IR relations in AdS dynamics,''
Phys.\ Rev.\ D {\bf 59}, 065011 (1999).
[hep-th/9809022].
}

\lref\MinwallaXI{
  S.~Minwalla and N.~Seiberg,
  ``Comments on the IIA (NS)five-brane,''
JHEP {\bf 9906}, 007 (1999).
[hep-th/9904142].
}

\lref\BerensteinGJ{
  D.~Berenstein and R.~G.~Leigh,
  ``Space-time supersymmetry in AdS(3) backgrounds,''
Phys.\ Lett.\ B {\bf 458}, 297 (1999).
[hep-th/9904040].
}

\lref\gik{Work in progress.}

\lref\GiveonZM{
  A.~Giveon, D.~Kutasov and O.~Pelc,
  ``Holography for noncritical superstrings,''
JHEP {\bf 9910}, 035 (1999).
[hep-th/9907178].
}

\lref\IsraelRY{
  D.~Israel, C.~Kounnas and M.~P.~Petropoulos,
  ``Superstrings on NS5 backgrounds, deformed AdS(3) and holography,''
JHEP {\bf 0310}, 028 (2003).
[hep-th/0306053].
}

\lref\ForsteWP{
  S.~Forste,
  ``A Truly marginal deformation of SL(2, R) in a null direction,''
Phys.\ Lett.\ B {\bf 338}, 36 (1994).
[hep-th/9407198].
}

\lref\GiveonGB{
  A.~Giveon, E.~Rabinovici and A.~Sever,
  ``Strings in singular time dependent backgrounds,''
Fortsch.\ Phys.\  {\bf 51}, 805 (2003).
[hep-th/0305137].
}

\lref\AharonyVK{
  O.~Aharony, B.~Fiol, D.~Kutasov and D.~A.~Sahakyan,
  ``Little string theory and heterotic / type II duality,''
Nucl.\ Phys.\ B {\bf 679}, 3 (2004).
[hep-th/0310197].
}

\lref\BerkoozUG{
  M.~Berkooz, A.~Sever and A.~Shomer,
  ``'Double trace' deformations, boundary conditions and space-time singularities,''
JHEP {\bf 0205}, 034 (2002).
[hep-th/0112264].
}

\lref\WittenUA{
  E.~Witten,
  ``Multitrace operators, boundary conditions, and AdS / CFT correspondence,''
[hep-th/0112258].
}

\lref\MaldacenaUZ{
  J.~M.~Maldacena, J.~Michelson and A.~Strominger,
  ``Anti-de Sitter fragmentation,''
JHEP {\bf 9902}, 011 (1999).
[hep-th/9812073].
}

\lref\SeibergXZ{
  N.~Seiberg and E.~Witten,
  ``The D1 / D5 system and singular CFT,''
JHEP {\bf 9904}, 017 (1999).
[hep-th/9903224].
}

\lref\AharonyXN{
  O.~Aharony, A.~Giveon and D.~Kutasov,
  ``LSZ in LST,''
Nucl.\ Phys.\ B {\bf 691}, 3 (2004).
[hep-th/0404016].
}

\lref\GiveonPX{
  A.~Giveon and D.~Kutasov,
  ``Little string theory in a double scaling limit,''
JHEP {\bf 9910}, 034 (1999).
[hep-th/9909110].
}

\lref\GiveonTQ{
  A.~Giveon and D.~Kutasov,
  ``Comments on double scaled little string theory,''
JHEP {\bf 0001}, 023 (2000).
[hep-th/9911039].
}

\lref\AharonyUB{
  O.~Aharony, M.~Berkooz, D.~Kutasov and N.~Seiberg,
  ``Linear dilatons, NS five-branes and holography,''
JHEP {\bf 9810}, 004 (1998).
[hep-th/9808149].
}

\lref\HorneGN{
  J.~H.~Horne and G.~T.~Horowitz,
  ``Exact black string solutions in three-dimensions,''
Nucl.\ Phys.\ B {\bf 368}, 444 (1992).
[hep-th/9108001].
}

\lref\HorowitzEI{
  G.~T.~Horowitz and A.~A.~Tseytlin,
  ``On exact solutions and singularities in string theory,''
Phys.\ Rev.\ D {\bf 50}, 5204 (1994).
[hep-th/9406067].
}

\lref\GiveonMYJ{
  A.~Giveon, N.~Itzhaki and D.~Kutasov,
  ``A Solvable Irrelevant Deformation of $AdS_3/CFT_2$,''
[arXiv:1707.05800 [hep-th]].
}

\lref\WeinbergMT{
  S.~Weinberg,
  ``The Quantum theory of fields. Vol. 1: Foundations,'' Chapter 10.
}

\lref\ItzhakiZR{
  N.~Itzhaki, D.~Kutasov and N.~Seiberg,
  ``Non-supersymmetric deformations of non-critical superstrings,''
JHEP {\bf 0512}, 035 (2005).
[hep-th/0510087].
}

\lref\MartinecZTD{
  E.~J.~Martinec and S.~Massai,
  ``String Theory of Supertubes,''
[arXiv:1705.10844 [hep-th]].
}

\lref\MaldacenaKM{
  J.~M.~Maldacena and H.~Ooguri,
  ``Strings in AdS(3) and the SL(2,R) WZW model. Part 3. Correlation functions,''
Phys.\ Rev.\ D {\bf 65}, 106006 (2002).
[hep-th/0111180].
}

\lref\CallanAT{
  C.~G.~Callan, Jr., J.~A.~Harvey and A.~Strominger,
  ``Supersymmetric string solitons,''
In *Trieste 1991, Proceedings, String theory and quantum gravity '91* 208-244 and Chicago Univ. - EFI 91-066 (91/11,rec.Feb.92) 42 p.
[hep-th/9112030].
}

\lref\KutasovXB{
  D.~Kutasov,
  ``Geometry on the Space of Conformal Field Theories and Contact Terms,''
Phys.\ Lett.\ B {\bf 220}, 153 (1989).
}

\lref\GiribetIMM{
  G.~Giribet,
  ``$T\bar{T}$-deformations, AdS/CFT and correlation functions,''
[arXiv:1711.02716 [hep-th]].
}

\Title{} {\centerline{Holography Beyond AdS}}

\bigskip
\centerline{\it Meseret Asrat${}^{1}$, Amit Giveon${}^{2}$, Nissan Itzhaki${}^{3}$ and David Kutasov${}^{1}$}
\bigskip
\smallskip
\centerline{${}^1$EFI and Department of Physics, University of
Chicago} \centerline{5640 S. Ellis Av., Chicago, IL 60637, USA }
\smallskip
\centerline{${}^{2}$Racah Institute of Physics, The Hebrew
University} \centerline{Jerusalem 91904, Israel}
\smallskip
\centerline{${}^{3}$ Physics Department, Tel-Aviv University, Israel} \centerline{Ramat-Aviv, 69978, Israel}

\smallskip

\vglue .3cm

\bigskip

\bigskip
\noindent

We continue our study of string theory in a background that interpolates between $AdS_3$ in the infrared and a linear dilaton spacetime $\IR^{1,1}\times\IR_\phi$ in the UV. This background corresponds via holography to a $CFT_2$ deformed by a certain irrelevant operator of dimension $(2,2)$. We show that for two point functions of local operators in the infrared CFT, conformal perturbation theory in this irrelevant operator has a finite radius of convergence in momentum space, and one can use it to flow up the renormalization group. The spectral density develops an imaginary part above a certain critical value of the spectral parameter; this appears to be related to the non-locality of the theory. In position space, conformal perturbation theory has a vanishing radius of convergence; the leading non-perturbative effect is an imaginary part of the two point function.

\bigskip

\Date{11/17}


\newsec{Introduction}

In two recent papers \refs{\GiveonNIE,\GiveonMYJ} we studied a string background that interpolates between a three dimensional linear dilaton background in the ultraviolet (UV) and $AdS_3$ in the infrared (IR).\foot{In both cases times a compact space that is a spectator under the deformation.} From the UV point of view, this background can be interpreted as the bulk description of Little String Theory (LST) in a vacuum with $N$ fundamental strings \refs{\AharonyUB,\GiveonZM}. From the IR point of view, it can be thought of as an irrelevant deformation of the two dimensional conformal field theory $(CFT_2)$ dual to the above $AdS_3$ background.

Apriori, one would expect the second point of view not to be useful for studying the theory, since it corresponds to flowing up the renormalization group (RG), a process that is in general highly ambiguous. However, as pointed out in  \refs{\GiveonNIE,\GiveonMYJ}, there are reasons to believe that in this case the situation is better:
\item{(1)} While from the point of view of the $CFT_2$, the (irrelevant) deformation in question is not expected to be under control, in the dual string theory on $AdS_3$ it corresponds to adding to the worldsheet Lagrangian a marginal operator, which moreover has a current-current (null abelian Thirring) form, and thus is expected to be well behaved (in fact the deformed worldsheet theory is exactly solvable).
\item{(2)} The deformation in question shares many properties with the $T\bar T$ deformation of  $CFT_2$, studied recently in \refs{\SmirnovLQW,\CavagliaODA}. The latter was argued to be well behaved and in fact the deformed theory appears to be exactly solvable.

\noindent
One of the interesting properties of the theories studied in \refs{\GiveonNIE,\GiveonMYJ} and \refs{\SmirnovLQW,\CavagliaODA} is that their spectrum approaches that of a $CFT_2$ in the IR (\ie\ the corresponding entropy goes like $S_{IR}\sim\sqrt E$), while in the UV one finds a Hagedorn spectrum $S_{UV}\sim E$. Thus, the short distance behavior of these theories is not governed by a UV fixed point -- they are not local QFT's. Hence, if one can study these theories by starting with the infrared $CFT_2$ and flowing up the RG, they provide an example of non-local theories that can be understood non-perturbatively.

A natural question is how the non-locality of these theories is reflected in the structure of correlation functions of operators that are local in the infrared CFT. In this note we will initiate the study of this problem in the model of \refs{\GiveonNIE,\GiveonMYJ}.
Since this model corresponds to a well behaved vacuum of weakly coupled string theory, we can calculate such correlation functions in an expansion in an effective string coupling (a $1/N$ expansion, or equivalently a large central charge expansion in the IR $CFT_2$), and study their properties. We are primarily interested in the answers to two questions:

\item{(a)} To what extent can we understand these correlation functions by flowing up the RG from the infrared fixed point?

\item{(b)} How does the non-locality of the deformed theory manifest  itself in the structure of these correlation functions?

\noindent
We will perform the calculations using the bulk description of the theory, but this is likely a technicality. If we understood the $CFT_2$ dual of the infrared $AdS_3$ background sufficiently well, we could perform the same calculations directly in the $CFT_2$, and due to the $AdS/CFT$ correspondence we would expect to get the same results.

A natural question is what properties do we expect the correlation functions of the deformed theory to have. According to \AharonyUB, the non-locality of the theory is expected to be reflected in the fact that correlation functions in momentum space are well behaved, but their position space counterparts are not. We will discuss to what extent these expectations are realized.

Note added: The correlation functions we study were also considered in a recent paper by G. Giribet \GiribetIMM. 

\newsec{The construction}

The starting point of our discussion is (type II) string theory in the background $AdS_3\times\NN$, where $\NN$ is a compact manifold whose precise properties will not play a role below.\foot{Well studied examples include $\NN=S^3\times T^4$ and  $S^3\times K_3$, which exhibit $(4,4)$ superconformal symmetry on the boundary, and $\NN=S^1\times\hat\NN$, with $\hat\NN$ a (worldsheet) $(2,2)$ superconformal background, which gives rise to (2,2) superconformal symmetry on the boundary. Supersymmetry will not play a role in our analysis below, but we expect that incorporating it into the discussion will lead to further insights.} As discussed in detail in \refs{\GiveonNIE,\GiveonMYJ}, the deformation we are interested in corresponds in the boundary $CFT_2$  to adding to the Lagrangian the term
\eqn\delddd{\delta\CL_{\rm b}=\lambda D(x),}
where $D(x)$ is a certain dimension $(2,2)$ quasi-primary of the spacetime conformal symmetry constructed in \KutasovXU. In supersymmetric examples, $D(x)$ is the top component of a superfield, so this deformation preserves SUSY while breaking conformal symmetry.

The above deformation can be described by adding to the worldsheet Lagrangian of string theory on $AdS_3$ the term
\eqn\deltall{\delta\CL_{\rm ws}=\lambda J^-\bar J^-,}
where $J^-$ is the worldsheet $SL(2,\IR)$ current whose zero mode gives rise to the spacetime (or boundary) Virasoro generator $L_{-1}$.

In the bulk description, $\delta\CL_{\rm ws}$ \deltall\ can be thought of as a supergravity deformation; the deformed geometry takes the form $\MM_3\times\NN$, where $\MM_3$ is described by the worldsheet Lagrangian \refs{\ForsteWP,\GiveonZM}
(see also \IsraelRY)
\eqn\wwssll{\LL=k\partial\phi\bar\partial\phi+{\lambda\over \lambda+e^{-2\phi}}\partial x^+\bar\partial x^-,
}
with a dilaton that goes like $\Phi\sim-\ln(1+\lambda e^{2\phi})$. The background \wwssll\ interpolates between a linear dilaton spacetime $\IR_\phi\times\IR^{1,1}$ in the UV region $\phi\to\infty$, and $AdS_3$ in the IR  $\phi\to-\infty$ (where it is natural to rescale $x^\pm$  by a factor of $\sqrt{\lambda/k}$, \GiveonMYJ). The coupling $\lambda$ sets the scale at which the transition takes place.

A useful description of the deformed model \deltall\ is obtained by starting with the background
\eqn\twelved{\IR^{1,1}\times AdS_3\times\NN,}
and gauging the null current
\eqn\nullcur{i\partial(y-t)+\epsilon J^-,}
and its right-moving analog $i\bar\partial(y+t)+\epsilon\bar J^-$. Here $(t,y)$ are (canonically normalized) coordinates on $\IR^{1,1}$, and $\epsilon$ is related to $\lambda$ as $\lambda\sim\epsilon^2$.

To define observables in the background \wwssll\  using the above coset description,  it is useful to recall the form of the observables in string theory on $AdS_3$ (see \eg\ \refs{\KutasovXU,\GiveonUP} for more detailed discussions and references).  A large class of such observables is given by vertex operators in the (NS,NS) sector, which take the form (in the $(-1,-1)$ picture)
\eqn\locobs{\hat\OO(x)=\int d^2z e^{-\varphi-\bar\varphi}\Phi_h(x;z)\OO(z).}
Here $\varphi$, $\bar\varphi$ are worldsheet fields associated with the superconformal ghosts, that keep track of the picture. $\Phi_h(x;z)$ are natural vertex operators on $AdS_3$, labeled by position on the boundary, $x$, and on the worldsheet, $z$. $\OO$ is an ($\NN=1$ superconformal primary) operator in the worldsheet theory on $\NN$. The operator \locobs\ satisfies the mass-shell condition
\eqn\massshell{-{h(h-1)\over k}+\Delta_\OO=\half~,}
which relates the scaling dimension of the operator $\hat\OO(x)$ in the spacetime (or boundary) CFT, $h$, to the worldsheet scaling dimension of the operator $\OO$, $\Delta_\OO$.\foot{We take $\bar\Delta_\OO=\Delta_\OO$, so that $\bar h=h$.}

The operator $\hat\OO(x)$ \locobs\ is a local operator in the boundary $CFT_2$. When we turn on the perturbation  \deltall, it is deformed to an operator in the perturbed theory corresponding to the background \wwssll, whose form was discussed in \GiveonMYJ. We next review this construction.

First, to facilitate the gauging, we need to Fourier transform the operators $\Phi_h(x;z)$ from the position $(x,\bar x)$ to the momentum $(p,\bar p)$ basis on the boundary. This gives rise to operators, which we will denote by $\Phi_h(p;z)$ (in a slight abuse of notation), which are eigenfunctions of the currents $(J^-,\bar J^-)$ with eigenvalues $(p,\bar p)$.  These operators behave like
\eqn\ppphhii{\Phi_h(p)= f_h(\phi)e^{i\vec p\cdot\vec\gamma},}
where the coordinates $\vec\gamma=(\gamma,\bar\gamma)$ parametrize the boundary of $AdS_3$, and $\phi$ parametrizes the radial direction. Near the boundary at $\phi\to\infty$, one has $f_h(\phi)\sim e^{\sqrt{2}(h-1)\phi}$. Gauge invariance implies that (the Fourier transforms of) the operators \locobs\ must be replaced in the deformed theory by
\eqn\observ{\hat\OO(p)=\int d^2z e^{-\varphi-\bar\varphi}\Phi_h(p)e^{-i(\omega t+p_yy)}\OO.}
The mass-shell condition \massshell\ is now deformed to
\eqn\mmaass{-{h(h-1)\over k}+{\alpha'\over4}(p_y^2-\omega^2)+\Delta_\OO=\half~.}
Moreover, gauge invariance sets $\omega=\epsilon p_0$ and $p_y=\epsilon p_1$, where $(p_0,p_1)$ are components of the vector $\vec p$ in \ppphhii.

The operators \observ\ are natural observables in the background \wwssll. In the IR limit $p_y,\omega\ll m_s$, they reduce to those in $AdS_3$ (the Fourier transforms of \locobs), and their correlation functions reduce to those of the $CFT_2$ dual to string theory in $AdS_3\times\NN$. To study these correlation functions away from the infrared limit, we need to compute them for general momenta.  In this note we will focus on the two point function, which we turn to in the next section.

\newsec{Two point function in momentum space}

The goal of this section is to compute the two point function of the operators \observ, $\langle \hat\OO(p)\hat\OO(-p)\rangle$. We will do this calculation in Euclidean space, \ie\ rotate $t\to i\tau$ and $\omega\to -ip_\tau$ in \observ, \mmaass. To do the calculation, we need to specify the normalizations of the operators in \observ. We will take the operator $\OO$ from the internal worldsheet CFT, $\NN$, to be normalized to one, $\langle \OO(z)\OO(w)\rangle=1/|z-w|^{4\Delta_\OO}$. This choice will not play an important role below, essentially because the internal CFT is a spectator under the deformation \deltall.

A more significant choice (as we will see below) is that of the normalization of the operator $\Phi_h$ in \observ. Recall that in conformal field theory on $AdS_3$ one has \TeschnerFT
\eqn\adsthreetwo{\langle\Phi_h(x;z)\Phi_{h'}(y;w)\rangle =  \delta(h - h') \frac{B(h)}{|z - w|^{4\Delta_h}|x - y|^{4h}}~,}
where $\Delta_h=-{h(h-1)\over k}$ is the dimension of $\Phi_h$ (see \massshell), and $B(h)$ is a function whose precise form depends on the normalization of $\Phi_h$. In a natural normalization, it takes the form
\eqn\formbh{B(h) = {k\over\pi}\nu^{2h-1}\gamma\left(1-{2h - 1\over k}\right),  \qquad \gamma(x) = {\Gamma(x)\over\Gamma(1 - x)},}
where $\nu$ is a constant whose value can be adjusted by shifting the radial coordinate $\phi$ in \wwssll\ or, equivalently, by a scale transformation in the boundary CFT \GiveonUP.

Consider first the undeformed theory, corresponding to the background $AdS_3\times\NN$. Using the worldsheet two point functions above, one can compute the two point function of the operators \locobs\ in the boundary CFT. One finds \refs{\GiveonUP,\MaldacenaKM}
\eqn\bndrytwo{\langle\hat\OO(x)\hat\OO(y)\rangle ={D(h)\over |x - y|^{4h}} , \qquad D(h) = (2h - 1)B(h).}
The calculation that led to \bndrytwo\ was done in a particular normalization of the operators $\Phi_h$, \adsthreetwo, \formbh. One could rescale the operators by an arbitrary smooth function of $h$, which would change the normalization $D(h)$. In $AdS_3$ such a rescaling is harmless, and does not change the essential physics. As we will see, this is the case in the deformed background \wwssll\ as well, although less trivially so.

As mentioned above, to study the deformed background $\MM_3\times\NN$ we need to Fourier transform the operators \locobs. The Fourier transform of the two point function \bndrytwo\ is
\eqn\momsp{ D(h) \int d^2x e^{i\vec{p}\cdot \vec{x}}|x|^{-4h}=\pi D(h)\gamma(1-2h)\left(\frac{p^2}{4}\right)^{2h - 1},}
where we used the notation $\vec p=(p_\tau, p_y)$. The momentum space two point function \momsp\ has poles when $2h-1=n\in Z_+$. These poles have a simple interpretation -- they correspond to cases where the two point function behaves like $p^{2n}\ln p^2$. One can analytically continue the two point function \momsp\ in the $h$ plane around these poles, and we will do this below.

The momentum space two point function \momsp, with $h$ given by the solution of the mass-shell condition \massshell, is valid in the undeformed background $AdS_3\times\NN$. However, looking back at \observ, we see that the same formula describes the two point function in the deformed background $\MM_3\times\NN$, except now $h$ is given by a solution of the deformed mass-shell condition \mmaass.

The dimensionful coupling $\lambda$, \deltall, does not appear in this mass-shell condition. One can think of this in the following  way. In \observ\ we parametrized the boundary in terms of the coordinates $(t,y)$, and their conjugate momenta $(\omega, p_y)$, or after Wick rotation $(\tau,y)$ and $(p_\tau, p_y)$. These coordinates are related by a factor of $\epsilon$, \nullcur, to the natural coordinates on the boundary of $AdS_3$, in terms of which the low energy $CFT_2$ is phrased, $(\gamma,\bar\gamma)$.

As explained in  \refs{\GiveonNIE,\GiveonMYJ}, one can choose a value for $\lambda$ so that the rescaling factor between the coordinates on the boundary of $AdS_3$ and linear dilaton spacetime is equal to one. In that case, the mass-shell condition \mmaass\ holds in terms of both. More generally, if we want to express the mass-shell condition in terms of the momenta conjugate to the coordinates on the boundary of $AdS_3$, $(\gamma,\bar\gamma)$, we need to replace $(\omega,p_y)$ by $\epsilon (p_0,p_1)$ in \mmaass, which gives rise to a factor of $\epsilon^2\sim\lambda$ in front of the $p^2$ term.

Since we will be interested in the dependence of the correlation functions on $\lambda$, we will find it convenient to exhibit the factors of $\lambda$ explicitly. Hence, we will write the mass-shell condition \mmaass\ in the form
\eqn\mmaassnew{-{1\over k}h_{p^2}(h_{p^2}-1)+\half\lambda p^2+\Delta_\OO=\half~,}
where $p^2=p_\tau^2+p_y^2$ (in Euclidean space). Denoting by $h$ the value of $h_{p^2}$ in the undeformed theory, \massshell, $h=\lim_{p\to 0}h_{p^2}$, one has
\eqn\xxx{ 2h_{p^2} - 1 = \sqrt{(2h - 1)^2 + 2\lambda k p^2}~.
}
One can think of \xxx\ as describing the dependence of the dimension of the operator $\hat O(p)$ \observ\ on the scale. This kind of dependence is standard in non-conformal  theories -- it reflects the fact that $\hat O$ has a non-vanishing anomalous dimension.

Combining all the elements of the above discussion, we find the two point function of the operators \observ,
\eqn\yyy{ \langle\hat\O(p)\hat\O(-p)\rangle  = \pi D(h_{p^2})\gamma(1-2h_{p^2})\left(\frac{p^2}{4}\right)^{2h_{p^2} - 1}.}
In the rest of this section we will discuss this result.

The first question we would like to address is the dependence of the result on the normalization of the operators \locobs\ in string theory on $AdS_3$. As we mentioned above, before we perturb the theory by \deltall, we can rescale these operators by any function of $h$ without changing the physics. However, \yyy\ seems to suggest that after the deformation the result does depend on this normalization, since now the factor $D(h_{p^2})$ depends on the momentum, and changing $D(h)$ changes the momentum dependence of the two point function.

One way to understand the origin of this dependence from the perspective of the boundary theory is the following. We start with the observable $\hat\OO(x)$ in the IR CFT, which has dimension $(h,h)$ and some particular normalization, \eg\ \bndrytwo. We then add to the Lagrangian of the boundary theory the irrelevant perturbation \delddd, and evaluate the correlation function \yyy\ in the perturbed theory, \eg\ by using conformal perturbation theory. This calculation is sensitive to the contact terms of the perturbing operator $D(x)$ \delddd\ with the operators $\hat\OO(x)$. Since $D(x)$ is a dimension $(2,2)$ operator, and $\lambda$ has dimension $(-1,-1)$, these contact terms take the form
\eqn\conterms{D(x)\OO(y)=\delta^2(x-y) F(\lambda\partial_y\partial_{\bar y})\partial_y\partial_{\bar y}\OO(y),}
where $F(z)=F_0+a_1z+a_2z^2+\cdots$ is an arbitrary smooth function of its argument. These contact terms give rise to a $\lambda$ (or, equivalently, momentum) dependent redefinition of the operator $\OO(x)$, and they are the origin of the dependence of the two point function  \yyy\ on the normalization in our bulk construction.

The situation is similar to the study of the geometry of the space of conformal field theories in  \KutasovXB. There, contact terms provide information about the connection on the space of theories, which is not reparametrization invariant information on this space, however the metric on moduli space, which does depend on the choice of contact terms, still contains
invariant information, such as the curvature of the space of theories. Here, contact terms parametrize the freedom to redefine the operators in a smooth way as we move around in the space of theories, but the two point function \yyy\ still contains invariant information that does not depend on this freedom.

Thus, the relevant question is what is the invariant information contained in the two point function \yyy? One issue is the analyticity properties of the two point function in momentum space. As is clear from the expression \yyy, there are three possible sources of singularities: (a) the two point function in $AdS_3$, \bndrytwo, $D(h)$; (b) the factor $\gamma(1-2h)$; (c) the function $h_{p^2}$ \xxx. We next discuss them in turn.

As mentioned before, in a natural normalization, the two point function $D(h)$ is given by \formbh, \bndrytwo. It has singularities when $2h-1\in kZ_+$, which give rise via \xxx\ to singularities in momentum space. These singularities are poles of the sort discussed in \AharonyXN, where they were refered to as bulk poles. As we also discussed, one can rescale the operators \observ\ so that the function $D(h)$ changes from \formbh, \bndrytwo. We will assume that such a rescaling does not introduce additional singularities.

The key point about the singularities of $D(h)$ is that they reflect physics present already in the undeformed theory, the CFT dual to string theory on $AdS_3\times\NN$; they are related to the continuum of long strings present in this theory. One can work in a regime in which the singularities of the two point function of the deformed theory due to this factor do not play a role, \eg\ by taking the level $k$ to be large and studying operators with $h_{p^2}$ of order one.

The factor $\gamma(1-2h)$ has singularities when $2h-1\in Z_+$ that were mentioned above. According to \AharonyXN, they have a similar interpretation, as bulk poles. From our perspective here, these poles have a kinematic origin -- they appear as a result of the Fourier transform \momsp. Thus, one expects them not to play an important role in studying the theory. We will make this more precise in the next section, when we show that the spectral density does not include this factor.

Thus, the only non-trivial singularities of the two point function \yyy\ are those associated with the behavior of the function $h_{p^2}$ \xxx\ in momentum space. In the complex $p^2$ plane, the function $h_{p^2}$ has a branch cut starting at $p^2=-(2h-1)^2/2\lambda k$. Thus, for real Euclidean momenta, the two point function is smooth everywhere.

In Minkowski spacetime, we have $p^2=p_y^2-p_t^2$, so the two point function is smooth for spacelike momenta, and has a branch cut in the timelike domain, starting  at the point discussed in the previous paragarph. Conformal perturbation theory gives rise in this case to a well defined two point function in a finite range of timelike momenta. This case involves some subtleties that we will leave to future work.

The position of the branch cut is at the point where the modes \observ\ go from being non-normalizable to delta function normalizable $(h_{p^2}=1/2)$. Thus, it is associated with the transition from physics dominated by the $AdS_3$ part of the geometry \wwssll\ to physics dominated by the UV linear dilaton region. Therefore, this branch cut is directly related to the non-locality of the theory.

Another interesting question is to what extent we can define the theory by doing conformal perturbation theory around the infrared fixed point, \ie\ perturbing in the interaction Lagrangian \delddd. We see from the form of the result \yyy\ that the expansion has a finite radius of convergence. To be precise, if we strip off the factor $D(h_{p^2})\gamma(1-2h_{p^2})$ in \yyy, for the reasons explained above, we are left with a series in $\lambda$ that has a finite radius of convergence, $|(2h-1)^2/2kp^2|$. We can compute the coefficients in this series by doing conformal perturbation theory; this defines the two point function for arbitrary spacelike momenta, and in a finite range of timelike momenta. In this sense, one can in this case flow up the RG.

As explained in \refs{\GiveonNIE,\GiveonMYJ} and mentioned above, the deformed theory \delddd, \deltall\ is not a local QFT. This must be reflected in the form of the two point functions at large Euclidean momenta. In a theory that is governed by a UV fixed point, two point functions of local operators go, in momentum space, like powers of $p$ (as in  \momsp). It is natural to ask what happens in our case. As mentioned above, this calculation is ambiguous due to the freedom of changing the contact terms \conterms\ in the boundary theory, or changing the function $D(h)$  \bndrytwo\ in the bulk description. Nevertheless, for a given choice, one can calculate the high momentum behavior. Below we describe the results for the natural definition from the bulk perspective \adsthreetwo\ -- \bndrytwo.

We will take the dimension of the operator in the infrared CFT, $h$ \xxx, to be of order one, and study the large momentum limit of the two point function \yyy. There are two natural regimes to consider. One is the regime where $k\gg h_{p^2}\gg 1$. In that regime the two point function behaves like
\eqn\twoasymone{ \frac{k^2 \csc\left(\pi\sqrt{k\lambda p^2}\right)}{\sqrt{k\lambda p^2}}  \left(k\lambda \Lambda^2\right)^{-\sqrt{k\lambda p^2}},}
where $\Lambda$ is an arbitrary scale, whose value is related to $\nu$ in \formbh. This function behaves like $\exp\left(-2c\sqrt{\lambda kp^2}\right)$ for some constant $c$ (away from the poles). 

The exponential behavior of \twoasymone\ is actually physically insignificant. We mentioned above the freedom of shifting the radial coordinate $\phi$, which has the effect of changing the constant $\nu$ in \formbh, and as a consequence the scale $\Lambda$ in \twoasymone. Performing such a shift in a general $n$ point function of operators of the form \observ\ rescales the amplitude by a factor that goes at large momenta like $\prod_{j=1}^n \exp\left(\alpha p_j\right)$. Thus, the exponential factor in \twoasymone\ is a feature of the particular normalization of the operators, and can be removed, uniformly in all correlation functions, by normalizing the operators appropriately.\foot{For early discussions of this exponential behavior see \refs{\PeetWN,\MinwallaXI}.}

A second regime that one may consider is the asymptotic high Euclidean momentum regime $h_{p^2}\gg k$, where one finds
\eqn\twoasymtwo{ k\csc\left(\pi\sqrt{k\lambda p^2}\right)\csc\left(\pi\sqrt{\frac{\lambda p^2}{k}}\right)\left(k\lambda \Lambda^2\right)^{-\sqrt{k\lambda p^2}}\left(\frac{\lambda p^2}{k}\right)^{-\sqrt{\frac{\lambda p^2}{k}}}.
}
The exponential factor in $p$ can be removed as before, but now the two point function behaves like $p^{-\alpha p}$, with $\alpha$ a constant that can be read off \twoasymtwo. Clearly, the behavior of the two point function in this regime differs significantly from that of a local QFT.

The above discussion is reminiscent of the results of  \GiveonCMA\ on the high momentum behavior of the scattering phase from Euclidean black holes in string theory. There, as here, this behavior comes in the bulk from classical string theory effects, and is invisible in supergravity. The difference between the two analyses is that \GiveonCMA\ considers delta function normalizable wave-functions (scattering states), while here we studied non-normalizable wave-functions, relevant for holographic correlation functions. It would be interesting to investigate the connection between the two further.

\newsec{Spectral density and two point function in position space}

\subsec{Spectral density}

The Kallen-Lehmann representation relates the momentum space two point function to the spectral density $\rho(\mu^2)$ via the relation
 \eqn\foup{\langle\hat\OO(p)\hat\O(-p)\rangle = \int_{0}^{\infty} d\mu^2 \frac{\rho(\mu)}{\mu^2 + p^2}~.
}
We can use the form of the two point function \yyy\ to compute the spectral density $\rho$. In fact, we will solve a more general problem; we will assume that the two point function takes the form
\eqn\twoptfn{
\langle\hat\OO(p)\hat\OO(-p)\rangle = G(\lambda p^2)\left(\frac{p^2}{4}\right)^{F(\lambda p^2)},
}
where $F$, $G$ are arbitrary functions, smooth near the origin, and compute $\rho$ \foup. In our case, one has
 \eqn\fgx{F(\lambda p^2) = 2h_{p^2} -1, \qquad G(\lambda p^2) =   \pi D(h_{p^2})\gamma(1 - 2h_{p^2}).
}
Since we assume that $F, G$ are smooth functions, we can Taylor expand \twoptfn\ in $\lambda p^2$,  and write it as
\eqn\powerser{
\langle\hat\O(p)\hat\O(-p)\rangle = \sum_{n = 0}^{\infty} \frac{a_n(\lambda p^2)^n}{n!};\qquad a_n=
\partial_{\alpha}^n \left.\left(\left(\frac{p^2}{4}\right)^{F(\alpha)}G(\alpha)\right)\right|_{\alpha = 0}.
}
As discussed in the previous section, the series \powerser\ has a finite radius of convergence.

We next show that the spectral density $\rho(\mu)$ takes in this case the form

 \eqn\powerrho{
\rho(\mu) = \sum_{n = 0}^{\infty} \frac{b_n(-\lambda \mu^2)^n}{n!};\qquad b_n=\partial_\beta^n\left.\left(\left(\frac{\mu^2}{4}\right)^{F(\beta)} H(\beta)\right)\right|_{\beta = 0},
}
where the function $H(\beta)$ needs to be computed. To confirm this ansatz and compute $H(\beta)$, we plug \powerrho\ into the definition \foup\ and compute the integral over $\mu$. Doing that we find that
\eqn\formhx{H(\beta) = -\frac{1}{\pi}G(\beta)\sin\pi F(\beta).
}
Plugging in the form of $F$ and $G$ for our case, \fgx, we have
\eqn\ffhh{H(\lambda\mu^2) = \pi\frac{D(h_{\mu^2})}{\Gamma^2(2h_{\mu^2})}~.}
Plugging \ffhh\ into \powerrho\ and summing the Taylor series, one finds
\eqn\rhox{\rho(\mu) = \pi\frac{D(h_{-\mu^2})}{\Gamma^2(2h_{-\mu^2})}\left(\frac{\mu^2}{4}\right)^{2h_{-\mu^2} - 1}.
}
We see that the spectral density takes a similar form to the momentum space two point function \twoptfn, with $h_{p^2}\to h_{-\mu^2}$. In particular, the radius of convergence of the perturbative series \powerrho\ is again finite, and given by\foot{To be precise, for the $D(h)$  given by \formbh, \bndrytwo, this is the radius of convergence of the series for $h\ll k$, when we do not have to worry about singularities of $D(h)$.} $\lambda\mu^2=(2h-1)^2/2k$.  Beyond that point, there is a branch cut and the spectral density develops an imaginary part.

This is at first sight puzzling, since unitarity implies that the spectral density must be real (and positive). It is possible that the imaginary part  of $\rho$ is related to the non-locality of the theory. Indeed, in a local QFT the reality of $\rho$ follows from the requirement that local operators commute for spacelike separations  (see \eg\ \WeinbergMT). The operators $\hat\O$ \locobs\ are indeed local in the infrared CFT, but in the full theory one does not expect them to be local. The imaginary part of $\rho$ might be a sign of this non-locality. This interpretation is reasonable since, as before, the branch cut we find occurs precisely at the threshold for creating the continuum of states living in the linear dilaton region. 

We now study the large $\lambda\mu^2$ behavior of the spectral density \rhox. For $k \gg \lambda\mu^2 \gg 1$ we find
\eqn\lur{\rho(\mu)\sim e^{-i\sqrt{k\lambda\mu^2}\ln k\lambda\Lambda^2}  \left(\frac{\lambda \mu^2}{k^3}\right)^{-\frac{1}{2}} e^{ \pi\sqrt{k \lambda  \mu^2}},
}
while for $\lambda\mu^2 \gg k$ we have
\eqn\lurwd{\rho(\mu)\sim e^{-i\sqrt{k \lambda \mu^2}\ln\left[\left(\frac{k}{\lambda \mu^2}\right)^{\frac{1}{k}}k\lambda \Lambda^2 \right]} \left(k \lambda  \mu^2\right)^{\frac{1}{2}}e^{\pi\sqrt{k \lambda \mu^2}}.
}
The exponential growth of $\rho$ with $\mu$ is likely related to the Hagedorn spectrum of the theory.

\subsec{Position space two point function}

Our next task is to compute the two point function in position space, which is obtained by taking the Fourier transform of \yyy, or equivalently \foup. Using the identity
\eqn\propf{
\int d^2p \frac{e^{i\vec{p}\cdot\vec{y}}}{p^2 + \mu^2} = 2\pi K_{0}\left(\mu |y|\right),
}
the Fourier transform of  \foup\ gives
\eqn\xent{\langle\hat{\O}(y)\hat{\O}(0)\rangle =2\pi \int_{0}^{\infty}d\mu^2 \rho\left(\mu\right)K_{0}\left(\mu|y|\right),
}
where $K_0$ is the modified Bessel function of the second kind.

Since the spectral density \rhox\ is complex for $\lambda \mu^2 > (2h - 1)^2/2k$, we see from \xent\ that the two point function in position space is complex.  The imaginary part of the two point function is given by
\eqn\x{{\rm Im}\langle\hat{\O}(y)\hat{\O}(0)\rangle =2\pi \int_{\mu_0}^{\infty}d\mu^2 {\rm Im}\rho\left(\mu\right)K_{0}\left(\mu|y|\right),
}
where $\mu_0^2 = \frac{(2h - 1)^2}{2\lambda k}$.

In the regime where $\mu_0|y| \gg 1$, using $\lim_{|z|\to \infty}K_0(z) \approx e^{-z}$, we find that the imaginary part of the two point function goes like
\eqn\x{{\rm Im}\langle\hat{\O}(y)\hat{\O}(0)\rangle \sim e^{-\sqrt{\frac{(2h - 1)^2}{2k}\frac{|y|^2}{\lambda}}}.
}
Thus, we see that the imaginary part of the two point function is non--perturbative in the natural dimensionless expansion parameter $\lambda/y^2$.

Since the leading non-perturbative effect goes like $e^{-{cy\over\sqrt{\lambda}}}$, one expects on general grounds a perturbative series of the form
\eqn\pertser{\langle\hat{\O}(y)\hat{\O}(0)\rangle=\frac{1}{|y|^{4h}}\sum_{n=0}^\infty c_n\left(\lambda\over y^2\right)^n,}
where the (real) coefficients $c_n$ have the large order behavior
\eqn\largord{c_n\sim n^\alpha e^{\beta n}(2n)!~,}
with some constants $\alpha,\beta$. One can verify that this is indeed the case in the following way. As mentioned above, the perturbative expansion of the spectral density, \powerrho,  \rhox, has a finite radius of convergence. Substituting this expansion into \xent, and using the fact that for large $n$ the integral over $\mu$ is dominated by a saddle point at large $\mu y$, leads to the result \largord.

\newsec{Discussion}

In this note we initiated the study of correlation functions in the model of \refs{\GiveonNIE,\GiveonMYJ}, that interpolates between a two dimensional CFT in the IR and a theory with a Hagedorn density of states (Little String Theory) in the UV. We saw that two point functions of a large class of scalar operators exhibit an interesting analytic structure that seems to be related to the non-local nature of the theory. There are many things that remain to be understood. Below we mention a few examples.

We saw that in momentum space, conformal perturbation theory in the coupling $\lambda$, \delddd, \deltall, has a finite radius of convergence, and Euclidean two point functions are well defined for arbitrary momenta. Thus, one can define these correlation functions by starting from the IR and flowing up the RG. In position space the expansion in $\lambda$ is asymptotic, and the leading non-perturbative effect appears to be an imaginary part of the  two point functions. It would be interesting to understand this non-perturbative effect better. 

Our discussion was restricted to the case where the coupling $\lambda$ is positive. For negative $\lambda$ the geometry \wwssll\ exhibits a naked singularity, and it is interesting to see what the consequences of that are in the correlation functions computed here. The momentum space two point function \yyy\ has for negative $\lambda$ similar properties to those of the spectral density \rhox\ for positive $\lambda$ -- it is given by a perturbative expansion with a finite radius of convergence, and there is a branch cut starting at the place where $h_{p^2}$ \xxx\ reaches the critical value $1/2$. Beyond that point, the momentum space two point function develops an imaginary part. On the other hand,  the spectral density \rhox\ is now well defined for all finite $\mu$. It would be interesting to see if one can use these correlation functions to learn something about the physics associated with the naked singularity in the bulk geometry.

In this note we considered the simplest correlation functions -- two point functions of scalar operators that in the bulk description correspond to vertex operators of the form \locobs, \observ. There are many interesting generalizations of this analysis. One can consider higher point functions of these operators, that will probably reveal some additional properties of these theories. Another interesting generalization is to correlation functions of conserved currents. In the infrared CFT one typically has a symmetry algebra generated by holomorphic and anti-holomorphic currents, including the Virasoro generators $T(x)$, $\bar T(\bar x)$, affine Lie algebra generators $K^a(x)$ corresponding to various groups, and supercurrents. After the deformation, we expect these currents to give rise to conserved but non-holomorphic currents, such as the conserved stress-tensor $T_{\mu\nu}$. It would be interesting to calculate the two point functions of such currents, and use them to compute the Zamolodchikov $c$ function and other related functions.

Our considerations did not use supersymmetry, but most (perhaps all) consistent examples of the theories we discussed are supersymmetric. It would be interesting to bring the techniques of supersymmetric field theory to bear on the analysis of these theories.

As mentioned in the introduction, part of the motivation for our work, here and in \refs{\GiveonNIE,\GiveonMYJ}, is the results of \refs{\SmirnovLQW,\CavagliaODA} on $T\bar T$ deformed $CFT_2$. It would be interesting to calculate correlation functions analogous to those considered here for that case. There are reasons to believe that these correlation functions have a lot in common with those computed here. In particular, it would interesting to see if the momentum space two point functions are well defined for arbitrary momenta, to see if conformal perturbation theory has a finite radius of convergence in momentum space and zero radius of convergence in position space, and whether the position space two point function has a non-perturbative imaginary part, as in our case.

Another setting in string theory where irrelevant deformations were analyzed and found to be under some control is two dimensional string theory. The authors of \refs{\MooreGA,\HsuCM} analyzed the $c=1$ CFT coupled to two dimensional gravity, in the presence of a deformation of the form $\delta\LL=\lambda\cos pX$. Before coupling to gravity, the coupling $\lambda$ is relevant for $p$ smaller than some critical value (for which the dimension of $\cos pX$ is equal to 2), and irrelevant otherwise. When it is irrelevant, one does not expect to be able to flow up the RG generated by the coupling $\lambda$ in a unique way. After coupling to gravity, the coupling $\lambda$ becomes marginal, and one can compute correlation functions in the model in conformal perturbation theory. This was done, using the dual matrix model, in the above papers, and it was found that the resulting expansion is asymptotic in a fixed area representation, but at fixed cosmological constant the expansion has a finite radius of convergence. In \refs{\MooreGA,\HsuCM} it was not clear what the UV behavior of the model is in the case where the coupling $\lambda$ is irrelevant, and whether the model makes sense for all scales in this case. In our case one expects the model to exist for all scales, and the challenge is to understand the way its behavior is reflected in the form of the correlation functions. It would be interesting if the two models shed some light on each other.

\bigskip\bigskip
\noindent{\bf Acknowledgements:}
We thank O. Aharony, T. Banks, M. Berkooz, T. Dumitrescu, G. Giribet, D. Jafferis, I. Klebanov, J. Maldacena, S. Sethi, K. Skenderis and M. Strassler for discussions. The work of AG and NI is supported in part by the I-CORE Program of the Planning and Budgeting Committee and the Israel Science Foundation (Center No. 1937/12), and by a center of excellence supported by the Israel Science Foundation (grant number 1989/14). MA and DK are supported in part by DOE grant DE-SC0009924. DK thanks Tel Aviv University and the Hebrew University for hospitality during part of this work.

\listrefs
\end

------------------------------------

In this note we continue our  study \GiveonNIE\ of a certain deformation of string theory on $AdS_3$. This study was motivated
by two recent papers \refs{\SmirnovLQW,\CavagliaODA}, which argued that perturbing a two dimensional conformal field theory ($CFT_2$)  by a particular dimension $(2,2)$ operator, which behaves near the original CFT like the product of the holomorphic and anti-holomorphic components of the stress tensor, $T\bar T$, leads to a well defined theory, despite the fact that it corresponds to a flow up the renormalization group (RG). Moreover, the authors of these papers argued that the model is in a certain sense exactly solvable, and in particular computed its spectrum on $\IR\times S^1$. An interesting property of the resulting spectrum is that it smoothly interpolates between an entropy associated with a $CFT_2$ in the IR, and one that exhibits Hagedorn growth in the UV \GiveonNIE.

In the context of holography, the irrelevant deformation studied in \refs{\SmirnovLQW,\CavagliaODA} is a double trace deformation, which corresponds to a change of the boundary conditions of the bulk fields on $AdS_3$ \refs{\BerkoozUG,\WittenUA}. In \GiveonNIE, we pointed out that there is a single trace deformation of string theory on $AdS_3$ that shares many elements with that of \refs{\SmirnovLQW,\CavagliaODA}, but may be more interesting, since it modifies the local geometry of the bulk theory. Some of the features the two deformations have in common are:
\item{(1)} The perturbing operator is a quasi-primary of the (boundary, or spacetime) Virasoro algebra with dimension $(2,2)$. Moreover, the OPE of the perturbing operator with the stress tensor has the same structure in the two cases.
\item{(2)} The construction of \refs{\SmirnovLQW,\CavagliaODA} is universal, in the sense that all $CFT_2$'s contain the operator $T\bar T$ that drives the RG flow. Similarly, the construction of \GiveonNIE\ is universal, in the sense that the single trace operator that drives the RG flow exists in all vacua of string theory on $AdS_3$.
\item{(3)} In the string theory construction, the irrelevant deformation of the spacetime theory corresponds to a marginal deformation of the worldsheet one. Therefore, from the string theory perspective it is natural that the resulting spacetime theory is well defined, as in \refs{\SmirnovLQW,\CavagliaODA}.
\item{(4)} The marginal worldsheet deformation is by an operator bilinear in worldsheet currents, and as such is exactly solvable, as in \refs{\SmirnovLQW,\CavagliaODA}. In fact, as mentioned in \GiveonNIE\ and will be further discussed below, the deformed worldsheet theory can be thought of as a coset CFT, and one can use current algebra techniques to study it.
\item{(5)} The string theory construction of \GiveonNIE\ gives rise to a theory that interpolates between a $CFT_2$ entropy in the IR and a Hagedorn entropy in the UV, like in \refs{\SmirnovLQW,\CavagliaODA}.

Despite the close analogy between the two constructions, the precise relation between them is unclear, primarily due to our limited understanding of the spacetime CFT corresponding to string theory on $AdS_3$. In \GiveonNIE, it was pointed out that if we assume that the spacetime CFT takes the symmetric product form $\MM^p/S_p$, where $\MM$ is a CFT with central charge $6k$, and  $k$ is the level of the worldsheet $SL(2,\IR)$ current algebra in string theory on $AdS_3$, as suggested in \refs{\ArgurioTB,\GiveonCGS},
the string theory single trace deformation corresponds to a $T\bar T$ deformation of the block $\MM$. The high energy behavior of the entropy of the deformed symmetric product CFT was shown to agree with the Bekenstein-Hawking entropy of black holes in the deformed geometry induced by the single trace deformation.

In this note, we would like to comment on a few aspects of the construction of \GiveonNIE. In section 2, we describe this construction in terms of a coset CFT, which involves null gauging of a $10+2$ dimensional background. We comment briefly on observables in the theory, which are naturally described in terms of this coset CFT, and use it (in section 3) to describe the spectrum of states of the resulting model on a spatial circle. We show that for the superstring, in a particular vacuum with supersymmetry preserving boundary conditions on the circle, the spectrum one gets is the same as that of \refs{\SmirnovLQW,\CavagliaODA}, assuming the $\MM^p/S_p$ structure mentioned above. In section 4, we discuss our results and their relation to those of \GiveonMI\ on the string/black hole transition in $AdS_3$ and linear dilaton backgrounds.

\newsec{Coset description}

A large class of $(2,2)$ supersymmetric vacua of string theory on $AdS_3$ is obtained by studying the worldsheet theory on $AdS_3\times S^1\times \NN$, where $\NN$ is a compact background described by a $(2,2)$ superconformal worldsheet theory
(see \eg\ \refs{\GiveonNS,\GiveonJG,\BerensteinGJ,\ArgurioTB}). Spacetime SUSY leads to a chiral GSO projection, which acts as an orbifold on this background. A useful way of thinking about these backgrounds is as describing systems of $NS5$-branes wrapped around various surfaces in a way that preserves some supersymmetry, in a state with a large number of fundamental strings bound to the fivebranes \refs{\GiveonZM,\ArgurioTB}.

A special case of this construction, which is sufficient for our purposes, is the background corresponding to $k$ $NS$ fivebranes wrapped around a four manifold $\MM^4(=T^4$ or $K_3)$, and $p$ strings,
\eqn\aaa{AdS_3\times S^3\times \MM^4.}
As in \GiveonNIE, we are interested in deforming this background by adding to the worldsheet Lagrangian the term
\eqn\deltall{\delta\CL=\lambda J^-\bar J^-,}
where $J^-$ is the worldsheet $SL(2,\IR)$ current whose zero mode gives rise to the spacetime Virasoro generator $L_{-1}$. As described in \refs{\ForsteWP,\GiveonZM,\IsraelRY}, this marginal worldsheet deformation leads to an asymptotically linear dilaton geometry, which interpolates between the $(AdS_3)$ near-horizon geometry of both the strings and the fivebranes in the IR, and the linear dilaton (CHS \CallanAT) geometry of just the fivebranes in the UV.

To describe the deformed CFT as a coset, we start with the following $10+2$ dimensional background:\foot{Or, in the more general class of vacua mentioned above, $\IR^{1,1}\times AdS_3\times S^1\times \NN$.}
\eqn\twelved{\IR^{1,1}\times AdS_3\times S^3\times \MM^4.}
Later, when studying states on the cylinder, we will compactify the spatial direction in $\IR^{1,1}$ on a circle. The uncompactified geometry is useful for studying off-shell correlation functions, as in \refs{\AharonyVK,\AharonyXN}.

We note in passing that the background \twelved\ plays an important role in many studies of fivebranes in string theory. For example, the system of fivebranes on a circle \refs{\GiveonPX,\GiveonTQ}, known as Double Scaled Little String Theory (DSLST), involves the coset of \twelved\ by the null current $J^3-K^3$, where $J^3$ is the timelike $U(1)$ in $AdS_3$ and $K^3$ is a CSA generator of $SU(2)$ \IsraelIR. Systems of fivebranes in motion are described by modifying the null current to involve the time translation generator \ItzhakiZR. And, recently it has been shown \MartinecZTD\ that some of the Ramond ground states of the string-fivebrane system can be described by adding the null translation generator in $\IR^{1,1}$ (more precisely $\IR\times S^1$) to the null generator $J^3-K^3$ mentioned above. Other closely related cosets give rise to black holes
(see e.g. \refs{\HorneGN,\HorowitzEI} and appendix C of \GiveonMI)
and cosmological backgrounds (see e.g. \GiveonGB\ for a review).

To describe the construction of \GiveonNIE\ as a coset CFT, we gauge the null current
\eqn\nullcur{i\partial(y-t)+\epsilon J^-,}
where $(t,y)$ are coordinates on $\IR^{1,1}$ and, again, $y$ may be compact. The current \nullcur\ is null and thus anomaly free. We can also gauge the right-moving current $i\bar\partial(y+t)+\epsilon\bar J^-$.

To understand the geometry that we get by gauging \nullcur\ and its right-moving analog in \twelved, we start with the sigma model on $AdS_3\times \IR^{1,1}$, which is described by the worldsheet Lagrangian
\eqn\wslag{ \LL=k(\partial\phi\bar\partial\phi+e^{2\phi}\bar\partial\gamma\partial\bar\gamma)+\partial x^+\bar\partial x^-.
}
The coordinates $\gamma=\gamma^1-\gamma^0$, $\bar\gamma=\gamma^1+\gamma^0$ parametrize the boundary of $AdS_3$; $x^\pm=y\pm t$ are coordinates on $\IR^{1,1}$. The symmetry we would like to mod out by is
\eqn\gaugesym{\eqalign{x^-\to &\,\,x^-+\alpha\;;\qquad\gamma\to\gamma+\epsilon\alpha,\cr
x^+\to &\,\,x^++\bar\alpha\;;\qquad\bar\gamma\to\bar\gamma+\epsilon\bar\alpha,
}}
where $\alpha$, $\bar\alpha$ are the gauge parameters of the two null $U(1)$'s.\foot{In \gaugesym\ we chose an axial gauging. One could also perform a vector gauging, for which  $\bar\gamma\to\bar\gamma-\epsilon\bar\alpha$. This gives rise to a singular geometry \GiveonNIE.}

To implement the gauging, we modify  \wslag\ as follows:
\eqn\gaugedlag{ \LL=k\left[\partial\phi\bar\partial\phi+e^{2\phi}(\bar\partial\gamma+\epsilon\bar A)(\partial\bar\gamma+\epsilon A)\right]+(\partial x^++A)(\bar\partial x^-+\bar A).
}
Eliminating the gauge fields gives rise to the background
\eqn\wwssll{\LL=k\partial\phi\bar\partial\phi+{k\over k\epsilon^2+e^{-2\phi}}\bar\partial(\gamma-\epsilon x^-)\partial(\bar\gamma-\epsilon x^+),
}
with a dilaton that goes like $\Phi\sim-\ln(1+k\epsilon^2e^{2\phi})$. The metric, $B$ field and dilaton depend on the gauge invariant coordinates $\phi$, $\gamma^0-\epsilon t$ and $\gamma^1-\epsilon y$. We can fix the gauge $x^\pm=0$, which is natural in the infrared region $\phi\to-\infty$, or $\gamma=\bar\gamma=0$, which is natural near the boundary $\phi\to+\infty$. This gives rise to the well known geometry of strings and fivebranes (see \eg\ appendix A of \GiveonZM).

The parameter $\epsilon$ in \wwssll\ controls the transition from the near-horizon region of both the strings and the fivebranes $(e^{-\phi}\gg\epsilon\sqrt k)$, and the region where we are in the near horizon of the fivebranes but not of the strings $(e^{-\phi}\ll\epsilon\sqrt k)$. We can set it to any particular value by shifting $\phi$ and rescaling $(\gamma,\bar\gamma)$. The role of this parameter in the bulk theory is very similar to that of the coefficient of the irrelevant operator  in the Lagrangian of the corresponding boundary theory. The latter determines the scale at which the theory transitions from being dominated by the IR CFT, and the UV (Hagedorn) regime.

As mentioned above, the coset perspective is useful for studying correlation functions of off-shell operators in the theory. We will postpone a detailed discussion of these correlation functions to another publication, limiting our discussion here to a few comments.

Setting the deformation parameter $\lambda$ in \deltall\ to zero (or, equivalently, setting $\epsilon=0$ in the coset \nullcur), off-shell observables correspond to local operators on the boundary of $AdS_3$. A large class of such observables is given by vertex operators in the (NS,NS) sector, which take the form (in the $(-1,-1)$ picture)
\eqn\locobs{\hat\OO(x)=\int d^2z e^{-\varphi-\bar\varphi}\Phi_h(x;z)\OO(z).}
Here $\varphi$, $\bar\varphi$ are worldsheet fields associated with the superconformal ghosts, that keep track of the picture. $\Phi_h(x;z)$ are natural vertex operators on $AdS_3$, labeled by position on the boundary, $x$, and on the worldsheet, $z$ (see \eg\ \refs{\KutasovXU,\GiveonUP} for more detailed discussions and references), and $\OO$ is an ($\NN=1$ superconformal primary) operator in the worldsheet theory on $S^3\times \MM^4$, or more generally $S^1\times \NN$. The operator \locobs\ satisfies the mass-shell condition
\eqn\massshell{-{h(h-1)\over k}+\Delta_\OO=\half~,}
which relates the scaling dimension of the operator $\hat\OO(x)$ in the spacetime (or boundary) CFT, $h$, to the worldsheet scaling dimension of the operator $\OO$, $\Delta_\OO$.

When we add to the theory the $\IR^{1,1}$ factor in  \twelved\ and gauge the symmetry \gaugesym, the observables change as follows. First, to facilitate the gauging, we need to Fourier transform the operators $\Phi_h(x;z)$ from the position $(x)$ to the momentum $(p)$ basis on the boundary. This gives rise to operators, which we will denote by $\Phi_h(p;z)$ (in a slight abuse of notation), which are eigenfunctions of the currents $(J^-,\bar J^-)$ with eigenvalues $(p,\bar p)$.  These operators behave like
\eqn\ppphhii{\Phi_h(p)= f_h(\phi)e^{i\vec p\cdot\vec\gamma}.}
Near the boundary at $\phi\to\infty$, one has $f_h(\phi)\sim e^{\beta\phi}$, with $\beta$  proportional to $h-1$.
Gauge invariance implies that the operators \locobs\ must be replaced in the deformed theory by
\eqn\observ{\hat\OO(p)=\int d^2z e^{-\varphi-\bar\varphi}\Phi_h(p)e^{-i(\omega t+p_yy)}\OO.}
The mass-shell condition \massshell\ is now deformed to
\eqn\mmaass{-{h(h-1)\over k}+{\alpha'\over4}(p_y^2-\omega^2)+\Delta_\OO=\half~.}
Moreover, gauge invariance sets $\omega=\epsilon p_0$ and $p_y=\epsilon p_1$. Observables corresponding to non-normalizable vertex operators, \observ\ with $\half<h\in\IR$, are labeled by two-dimensional momentum $(\omega, p_y)$, with $h$ fixed by the mass-shell condition \mmaass. One can use the coset description to calculate off-shell correlation functions of such observables, and use them to study the high (and low) energy behavior of the theory. Note that the observables \observ\ are labeled by their momenta. One does not expect to be able to Fourier transform them to position space, due to the non-locality of the theory. This is believed to be a general feature of all vacua of Little String Theory, such as DSLST \refs{\GiveonPX,\GiveonTQ,\AharonyXN}.

\newsec{Comments on the spectrum}

To study the spectrum of the theory, we would like to compactify the spatial direction on the boundary of the geometry \wwssll\ on a circle. In the undeformed theory (\ie\ for $\epsilon=0$), we can do this by identifying $\gamma_1\sim \gamma_1+2\pi R_1$, with all fields satisfying periodic boundary conditions on the circle. This gives rise to the $M=J=0$ BTZ black hole geometry, which describes a Ramond-Ramond ground state of the boundary CFT.

The spectrum of perturbative string states in this background is continuous. This is easy to understand from the spacetime point of view. The background \aaa\ is obtained by adding to a linear dilaton background, of the form $\IR_\phi\times \IR_t\times S^1\times \MM^4\times S^3$, $p$ fundamental strings wrapping the $S^1$ \GiveonZM. The resulting state is BPS -- the strings preserve some of the supersymmetry of the original background. Thus, these strings do not feel a force attracting them to the fivebranes, and their excitations form a continuum. This continuum is described by the vertex operators of long strings constructed in \refs{\MaldacenaHW,\ArgurioTB}.

The deformation \deltall\ extends the background from the near-horizon geometry of both the strings and the fivebranes to just that of the fivebranes. This extension does not change the fact that the strings experience a flat potential; hence, one expects to find a continuum of states corresponding to strings wound around the spatial circle in \wwssll, and having an arbitrary radial momentum in $\phi$ and oscillation level.

Such states can be described as follows using the coset description of the previous section. Consider for example the (NS,NS) sector vertex operators in eq. \observ.  To describe states carrying arbitrary momentum $n$ and winding $w$ around the $y$ circle, we replace the factor $e^{ip_yy}$ by
\eqn\plr{e^{ip_yy}\to e^{ip_L y_L+ip_Ry_R},}
with
\eqn\ppllrr{p_L={n\over R}+{wR\over\alpha'}\;;\;\;\; p_R={n\over R}-{wR\over\alpha'}~.}
States carrying real radial momentum correspond to
\eqn\formhhh{h=j+1=\half+is\;;\;\;\;s\in\IR.}
The mass-shell condition \mmaass\ now takes the form
\eqn\newmass{{\alpha'\over4}\omega^2={\alpha'\over4}p_L^2-{j(j+1)\over k}+\Delta_\OO-\half~,}
and a similar equation for the other worldsheet chirality,
\eqn\barnewmass{{\alpha'\over4}\omega^2={\alpha'\over4}p_R^2-{j(j+1)\over k}+\bar\Delta_\OO-\half~.}
Adding \newmass\ and \barnewmass\ gives
\eqn\finmass{\omega^2=\left(n\over R\right)^2+\left(wR\over\alpha'\right)^2+{2\over\alpha'}
\left(-{2j(j+1)\over k}+\Delta_\OO+\bar\Delta_\OO-1\right).
}
The difference of the two gives
\eqn\difftwo{\bar\Delta_\OO-\Delta_\OO=nw.}
For $w=1$, the mass-shell conditions \finmass, \difftwo\ describe a string winding once around the spatial circle on the boundary in a particular excitation state labeled by $\OO$ and with a particular radial momentum labeled by $s$ \formhhh. To rewrite it in a more suggestive form, it is useful to measure the energy of this state relative to the energy of a BPS string wrapping the circle (which corresponds to the supersymmetric vacuum), \ie\ write
\eqn\formomega{\omega=E+{R\over\alpha'}~.}
It is also useful to recall that in the $AdS_3$ limit of the background \wwssll, the last term in \finmass\ is related  to the value of $L_0$, $h_1$, for a long string with the same quantum numbers \refs{\MaldacenaHW,\ArgurioTB},
\eqn\recallh{\eqalign{h_1-{k\over 4}=&-{j(j+1)\over k}+\Delta_\OO-{1\over 2}~,\cr
\bar h_1-{k\over 4}=&-{j(j+1)\over k}+\bar\Delta_\OO-{1\over 2}~.}
}
Plugging \formomega, \recallh\ into \finmass, \difftwo, we find (for $w=1$) the mass-shell condition
\eqn\thuszamo{\left(E+{R\over\alpha'}\right)^2-\left({R\over\alpha'}\right)^2
={2\over\alpha'}\left(h_1+\bar h_1-{k\over 2}\right)+\left({n\over R}\right)^2,}
and $\bar h_1-h_1=n$. The mass-shell condition \thuszamo\ agrees precisely with what one would find for a state with the dimensions \recallh\ in a CFT $\MM$ of central charge $c_\MM=6k$, upon a deformation of the sort studied in \refs{\SmirnovLQW,\CavagliaODA}, $\delta\LL=-tT\bar T$. To determine $t$ it is useful to recall that the quantity $R$ in \refs{\SmirnovLQW,\CavagliaODA}, $R_{QFT}$, is in our language the circumference of the spatial coordinate on the boundary of $AdS_3$, $\gamma_1$, and is related to the circumeference of the $y$ coordinate at infinity, $2\pi R$, by a factor of $\epsilon$,{\foot{Due to the relation between $y$ and $\gamma_1$ mentioned above.} \ie\ $R_{QFT}=2\pi R\epsilon$. Similarly, the energy in these papers is related to the energy here by a factor of $1/\epsilon$. Taking all this into account, and comparing \thuszamo\ to the spectrum in \refs{\SmirnovLQW,\CavagliaODA}, we find $t=\pi\alpha'\epsilon^2$.

As mentioned above, the physics is independent of $\epsilon$, since one can change it by rescaling the coordinates $(\gamma,\bar\gamma)$ (which also rescales the radius $R$) and shifting $\phi$. A convenient value  is $\epsilon=1$, since for that value the coordinates $(\gamma^0,\gamma^1)$ are normalized in the same way as the asymptotic coordinates $(t,y)$; this is the choice made in \GiveonNIE. Note that the value  $t=\pi\alpha'$, obtained here from the perturbative string spectrum (for $\epsilon=1$), agrees with that found in \GiveonNIE\ from black hole thermodynamics.

One can think of the states \thuszamo\ as belonging to the untwisted sector of the orbifold $\MM^p/S_p$. States with $w>1$ belong to the $Z_w$ twisted sector of the orbifold. To see this, one proceeds as follows. The analog of \formomega\ for this case is
\eqn\formomegaw{\omega=E+{wR\over\alpha'}~.}
Plugging this and \recallh\ into \finmass\ gives
\eqn\honehone{\left(E+{wR\over\alpha'}\right)^2-\left({wR\over\alpha'}\right)^2
={2\over\alpha'}\left(h_1+\bar h_1-{k\over 2}\right)+\left({n_w\over wR}\right)^2~,}
with $n_w=wn$. Comparing \honehone\ to \thuszamo, we see that the spectrum of strings with winding $w$ is the same as that of a string singly wound around a circle with radius $wR$ and momentum $n_w$. This agrees with the spectrum in the $Z_w$ twisted sector of $\MM^w/Z_w$, with $Z_w$ acting via cyclic permutation on the $w$ copies of $\MM$. Note that in the IR limit, $R/l_s\to\infty$, \honehone\ reduces to well known results in string theory on $AdS_3$ \refs{\MaldacenaHW,\ArgurioTB}, such as, \GiveonMI,
\eqn\hwhw{h_w={h_1\over w}+{k\over 4}\left(w-{1\over w}\right),}
describing long strings winding $w$ times around the boundary circle.

To recapitulate, we have shown that states with $w>0$ in the background \wwssll\ agree with those found in the symmetric product CFT $\MM^p/S_p$, with $\MM$ deformed via a $T\bar T$ deformation. Next we show that string theory on this background has states that do not fit this description, however these states decouple in the infrared limit, and thus are not visible in the IR CFT.

Since states with winding $w$ correspond to the $Z_w$ twisted sector, it is natural to expect that states with $w\le 0$ are not captured by the symmetric orbifold. For $w=0$  \finmass\ takes the form
\eqn\wzero{(ER)^2={2R^2\over\alpha'}\left(-{2j(j+1)\over k}+\Delta_\OO+\bar\Delta_\OO-1\right),
}
where we also set the momentum $n=0$ for simplicity. Thus, the dimensionless energy $ER$ diverges in the limit $R^2/\alpha'\to\infty$, which means that the states \wzero\ are not present in the undeformed CFT dual to string theory on $AdS_3$. In the language of the $T\bar T$ deformed theory, these are states with energy $E\sim1/\sqrt t$, which decouple in the IR limit $t\to 0$. States with $w<0$ decouple even faster when $t\to 0$, as their energy is bounded from below by ${2 |w| R \over \alpha'}$.

\newsec{Discussion}

In the previous section we described perturbative string excitations of the system of strings and fivebranes wrapping the circle labeled by $y$ (or $\gamma_1$) in \wwssll, with the fivebranes wrapping an additional four dimensional surface $\MM^4$ \twelved. We saw that the excitations of this system that correspond to one or more of the $p$ strings creating the background moving away from the fivebranes (while remaining in their near-horizon region) are well described by a dual boundary theory $\MM^p/S_p$, where $\MM$ is a CFT with central charge $c_\MM=6k$, which roughly corresponds to the theory of a single string. The worldsheet deformation \deltall, which corresponds to $\epsilon\not=0$ in \wwssll, is dual to a $T\bar T$ deformation (in the sense of \refs{\SmirnovLQW,\CavagliaODA}) of the CFT $\MM$.

There are also states which do not fit the $\MM^p/S_p$ structure, but they do not correspond to small excitations of the string/fivebrane system. States with $w<0$ can be thought of as obtained from the system of $p-1$ strings and $k$ fivebranes by adding to it an $F1-\bar{F1}$ pair, while states with $w=0$ correspond to adding to the string/fivebrane system an additional short string. In contrast, states with $w>0$ can be thought of as describing excitations of the $p$ strings forming the vacuum.

In this note, we focused on the spectrum of excitations of a particular Ramond-Ramond vacuum of the spacetime CFT, corresponding to the $M=J=0$ BTZ black hole. Modular invariance, spectral flow symmetry of the spacetime CFT, and explicit constructions imply the existence of many other Ramond and Neveu Schwartz vacua, and it would be interesting to extend our discussion to these vacua. We will postpone this to another publication.

Our results are related to those of \GiveonMI, which discussed the transition between perturbative strings and black holes in $AdS_3$ and linear dilaton backgrounds. The picture presented in that paper was the following. As one increases the available energy, first one creates perturbative long strings that can propagate towards the boundary. As the energy of these strings increases, they propagate to larger and larger radial distance, where the coupling of the theory on the strings grows, and eventually they cross over to black holes. In this sense, one can think of the long  strings as precursors of the black holes. Our results reinforce and extend this picture. The agreement found in  \GiveonNIE\ and here suggests that the symmetric product provides a good description of both the long strings and the black holes in $AdS_3$, and the deformation \deltall\ corresponds to the $T\bar T$ deformation in $\MM$.

Note that the above discussion is valid for the case where the level of the $SL(2,\IR)$ current algebra $k$ is larger than one. As discussed in detail in \GiveonMI, for $k<1$ the physics is different -- the black holes discussed in \GiveonMI\ are not normalizable, the coupling on the long strings in $AdS_3$ becomes weak near the boundary, and the generic high-energy states are these long strings. All this agrees with the $T\bar T$ deformed symmetric product theory, where in this case the spacetime CFT on $\MM$ is one in which the $SL(2,\IR)$ invariant vacuum is not in the spectrum.

and the last term in \finmass\ is thus related  to the value of $L_0$, $h_w$,
for $w$ long strings with the same quantum numbers, by
\refs{\MaldacenaHW,\ArgurioTB}
\eqn\recallhw{\eqalign{w\left(h_w-{kw\over 4}\right)=&-{j(j+1)\over k}+\Delta_\OO-{1\over 2}~,\cr
w\left(\bar h_w-{kw\over 4}\right)=&-{j(j+1)\over k}+\bar\Delta_\OO-{1\over 2}~.}
}
Plugging \formomegaw,\recallhw\ into \finmass,\difftwo, we find the mass-shell condition
\eqn\thuszamow{\left(E_w+{wR\over\alpha'}\right)^2-\left({wR\over\alpha'}\right)^2
={2w\over\alpha'}\left(h_w+\bar h_w-{kw\over 2}\right)+\left({n\over R}\right)^2,}
and $\bar h_w-h_w=n$. The mass-shell condition \thuszamow\ agrees apparently with what one would find for a state with the scaling dimensions $h_w-{c_w\over 24}$, $\bar h_w-{c_w\over 24}$, in a CFT of central charge $c_w=6kw$, upon a $t_wT\bar T$ deformation of the sort studied in \refs{\SmirnovLQW,\CavagliaODA}.
Moreover, \thuszamow\ can be rewritten as
\eqn\zamoww{\left({E}+{R\over\alpha'}\right)^2-\left({R\over\alpha'}\right)^2
={2\over\alpha'}\left({h}+{\bar h}-{k\over 2}\right)+\left({n/w\over R}\right)^2,}
where $\{E,h,\bar h\}=\{E_w,h_w,\bar h_w\}/w$,
which agrees precisely with what one would find for a state with the scaling dimensions
$h-{c_\MM\over 24}$, $\bar h-{c_\MM\over 24}$, in a CFT $\MM$ of central charge $c_\MM=6k$,
upon a $tT\bar T$ deformation to a state with energy ${E}$ and momentum ${n/w}$, of the sort studied in \refs{\SmirnovLQW,\CavagliaODA},
with a deformation parameter $t=\pi\alpha'$, that is identical to the one in (6.12) of \GiveonNIE,
and which is $w$ times bigger than the apparent one, $t_w={\pi\alpha'\over w}$, in \thuszamow.
Equations \thuszamow\ and \zamoww\ are compatible with the expectation that states in the $w$ winding sector
of the perturbative string spectrum around the $M=J=0$ BTZ black hole background,
correspond in the spacetime theory to states such as in the $w$-twisted sector of a symmetric orbifold $\MM^p/S_p$, with the deformation \deltall\ acting on the CFT $\MM$ as in \refs{\SmirnovLQW,\CavagliaODA}.

Comment:

Recall \GiveonMI\ that \recallhw\ implies that
\eqn\hwhw{h_w={h_1\over w}+{k\over 4}\left(w-{1\over w}\right)~.}
This is precisely the type of expression one finds in the $w$-twisted sector of a symmetric
orbifold, whose building block $\MM$ has $c_\MM=6k$.
In particular, plugging \hwhw\ into \thuszamow,
one finds
\eqn\honehone{\left(E_w+{wR\over\alpha'}\right)^2-\left({wR\over\alpha'}\right)^2
={2\over\alpha'}\left(h_1+\bar h_1-{k\over 2}\right)+\left({n_w\over wR}\right)^2~,}
which is the same as the equation for the $w=1$ sector, \thuszamo, with $R\to wR$, an energy $E_w$
above extremality, and momentum $n_w=wn$,
in harmony with a symmetric orbifold structure.